\documentclass[12pt]{article}   
\usepackage{amssymb}     
\usepackage{amsfonts}     
\usepackage{latexsym}       
\usepackage{epsfig}  
\usepackage{diagrams}
  
\def\endproof{$\Box$}   
\newtheorem{lm}{Lemma} \newcommand{\blm}{\begin{lm}\hspace{-5pt}{\bf .} \ }\newcommand{\elm}{\end{lm}}
\newcommand{\bit}{\begin{itemize}} \newcommand{\eit}{\end{itemize}\par\noindent}
\newcommand{\ben}{\begin{enumerate}} \newcommand{\een}{\end{enumerate}\par\noindent}
\newcommand{\beq}{\begin{equation}} \newcommand{\eeq}{\end{equation}\par\noindent}
\newcommand{\bla}{\left\{\begin{array}{l}} \newcommand{\ela}{\end{array}\right.}       
  
\begin{document}     
\noindent{{\Large{\bf Physical Traces\,:}}}

\smallskip\noindent
\noindent{{\Large{\bf Quantum vs.~Classical Information Processing}}}

\bigskip\par\noindent
{\normalsize{Samson Abramsky and Bob Coecke}}\footnote{Dusko Pavlovic, Phil Scott, Robert
Seely, Peter Selinger made inspiring remarks.}

\medskip\noindent 
{\footnotesize{University of Oxford, Oxford University Computing Laboratory\,,}}

\noindent 
{\footnotesize{Wolfson Building, Parks Road, Oxford, OX1 3QD, UK\,;}}

\noindent 
{\footnotesize{ e-mail: samson.abramsky/bob.coecke@comlab.ox.ac.uk\,.}}    

\bigskip\noindent 
{\footnotesize{Electronic Notes in Theoretical Computer Science {\bf
69} (2003).}}

\noindent
{\footnotesize{\tt www.elsevier.com/gej-ng/31/29/23/131/23/show/Products/notes/index.htt$\#$002}}
  
\par\bigskip\noindent  
{\bf Astract.}   
Within the Geometry of Interaction (GoI) paradigm, we present a setting that enables  qualitative
differences between classical and quantum processes to be explored.  The key construction is the physical
interpretation/realization of the traced monoidal categories of finite-dimensional vector spaces
with tensor product as monoidal structure and of finite sets and relations with Cartesian product as monoidal
structure, both of them providing a so-called wave-style GoI. The developments in this paper reveal that envisioning
state update due to quantum measurement as a process provides a
powerful tool for developing high-level approaches to
quantum information processing.   
   
\section{Introduction}  
 
Recall that a traced monoidal category is a symmetric monoidal category $(\mbox{\bf C},\otimes)$ such that
for every morphism $f:A\otimes C\to B\otimes C$ a trace $Tr_{A,B}^C(f):A\to B$ is specified and satisfies certain
axioms \cite{JSV}.  We refer to the available literature for explicit 
definitions, e.g. \cite{Abr,AHS,Hag,Sel2}.
These traced monoidal
categories play an important role in Geometry of Interaction (GoI) and 
game semantics, since 
every such traced monoidal category gives rise,  via the
GoI-construction, to a compact closed category in which 
composition corresponds to  interaction of strategies \cite{Abr}.  

Two qualitatively different families of GoI-style
categorical semantics for concurrent processes are identified in
\cite{Abr}, namely `particle-style' and `wave-style'
models, based on the interpretation of the tensor as categorical
coproduct and product respectively. The fundamentally different nature
of these families of models is further exposed in the elaboration in \cite{Sel1} of
Bainbridge's \cite{Bai} work on modelling flowcharts and networks, preceding the definition of traced monoidal
categories by two decades. 

The particle-style model is quite well understood and allows a ``physical
realization/interpretation'' in terms of a particle traveling through a network (see below).  The wave-style
models require the interpretation of the trace as a fixpoint, and hence 
require a domain-theoretic context. They have a reasonable
computational realization as dataflow networks, for which see
\cite{AJ}. Most mysterious are those examples in
which the tensor is neither categorical product or sum. The most
prominent examples are:
\begin{itemize}
\item The category of sets and relations
  with Cartesian product as the monoidal structure.
\item The category of finite-dimensional 
  vector spaces with the usual tensor product as the monoidal structure. 
\end{itemize}
These are both compact-closed categories, but their status in either
computational or physical terms is far from clear. 

In this paper, we
propose a physical realization for these categories, which we believe
can be the basis for some interesting new directions in quantum computation.
The main contributions of the present paper are as follows:
\bit
\item We delineate qualitative differences between computational properties of classical and quantum
systems, as incarnations of particle- and wave-style GoI-models respectively.
\item We realize the multiplicative fragment of linear logic by means
  of quantum devices, thus providing some new insights
into what is quantum about linear logic and what is linear about quantum logic. 
\item We develop a setting for quantum concurrency and quantum programming language development, implicitly
aiming towards high-level approaches to quantum algorithm design. 
\eit

Let us briefly sketch a particle-style model, namely the one that arises when considering the traced
monoidal category $(\mbox{\bf Rel},+)$ of relations with disjoint union  as tensor and with
``feedback'' as trace.  In this category, for
\[ R\subseteq (X+Z)\times (Y+Z) , \]
the trace is formally given by 
\[ x\,Tr_{X,Y}^Z(R)y\Leftrightarrow \exists z_1,
\ldots, z_n\in Z:xRz_1R \ldots R z_n Ry . \] 
The interpretation/realization of this category is as follows: We
envision a particle traveling trough a network where the objects $X$ specify the states $x\in X$ the particle
``can have'' at that stage (of traveling through the network) and the morphisms $R\subseteq X\times Y$ are
processes that impose a change of state from $x$ to a state in $\{y\in Y\mid xRy\}$, which is possibly empty
encoding that the process halts. The tensor
$X+Y$ is interpreted as disjoint union of  state sets.
$R+R'$ encodes parallel composition where,
depending on the initial state, either
$R$ or $R'$ will act on the particle.  The trace $Tr_{X,Y}^Z(R)$ encodes feedback, that is, entering in a state
$x\in X$ the particle will either halt, exit at $y\in Y$ or, exit at $z_1\in Z$ in which case it is fed back into
$R$ at the
$Z$ enterance, and so on, until it halts or exits at $y\in Y$.  

It is easily seen that such a ``one-dimensional''
perspective cannot hold for the following traced monoidal categories: 

\begin{itemize}
\item $(\mbox{\bf Rel},\times)$ of
relations with the Cartesian product as tensor and with as trace
\[ x\,Tr_{X,Y}^Z(R)y\Leftrightarrow \exists z\in
Z:(x,z)R(y,z) , \qquad
\] 
for 
$$R\subseteq (X\times Z)\times (Y\times Z).$$
\item  $(\mbox{\bf FDVec},\otimes)$ of
finite-dimensional vector spaces with the tensor product as tensor and  with as trace
$$
Tr_{{\cal V},{\cal W}}^{\cal U}(f):{\cal V}\to {\cal W}:v_i\mapsto\sum_{k,\alpha}f_{i\alpha{k\alpha}}
\cdot w_k 
$$   
for
$$f:{\cal V}\otimes {\cal U}\to{\cal W}\otimes {\cal U}:v_i\otimes e_j\mapsto\sum_{k,l}f_{ij{kl}}
\cdot (w_k\otimes e_{l})$$ 
where $\{v_i\}_i$, $\{w_k\}_k$, $\{e_l\}_l$ are the respective bases of 
${\cal V}$, ${\cal W}$ and ${\cal U}$. 
\end{itemize}

Let us now recall the core of the GoI construction as outlined in \cite{Abr}.
Given a traced monoidal category $(\mbox{\bf C},\otimes, Tr)$ we define a new category ${\cal
G}(\mbox{\bf C})$ with objects given by pairs $(A_+, A_-)$ of $\mbox{\bf C}$-objects.  A ${\cal G}(\mbox{\bf
C})$-morphism
\[ f:(A_+, A_-)\to(B_+, B_-) \] 
is a $\mbox{\bf C}$-morphism 
\[ \tilde{f}:A_+\otimes B_-\to A_-\otimes B_+ .  \]
Given
\[ f:(A_+, A_-)\to(B_+, B_-) , \qquad
g:(B_+, B_-)\to(C_+, C_-) , \] 
their composition 
\[ h=(f;g):(A_+, A_-)\to(C_+, C_-) \] 
in ${\cal G}(\mbox{\bf C})$ is
given by the trace 
\[ \tilde{h}:=Tr_{A_+\otimes C_-,C_+\otimes A_-}^{B_+\otimes
  B_-}(\tilde{f}\otimes \tilde{g}) . \] 
Via this construction one obtaines a compact closed category, that is, a $^*$-autonomous category with self-dual
tensor, where a $^*$-autonomous category is a symmetric monoidal closed category with a dualizing object. See for
example \cite{Hag} for an overview on these.  Another crucial property of this construction is that the 
category $\mbox{\bf C}$ embeds fully, faithfully and preserving the trace into ${\cal G}(\mbox{\bf C})$,
identifying
$A$ with 
$(A,I)$ where $I$ is the unit for the tensor. We thus embed the category $\mbox{\bf C}$ with ``sequential 
application'' as composition into the compact closed category ${\cal G}(\mbox{\bf C})$ with ``parallel
interaction'' as composition. 

The essence of the present paper lies in the construction
of an interaction category with true physical processes as morphisms
which compose in a concurrent fashion. Thus, even
though
$\mbox{\bf FDVec}$ is itself already compact closed, it makes sense to construct the ``larger'' compact closed
category ${\cal G}(\mbox{\bf FDVec})$. Moreover, the $^*$-operation
of the compact closed structure of $\mbox{\bf FDVec}$ has no direct physical implementation, but will
turn out to encode  an instance of the {\it duality between what measures and what is measured}.   
The construction for $\mbox{\bf FDVec}$ induces a realization/interpretation of the category $(\mbox{\bf
FRel},\times)$ of finite sets and relations with the Cartesian product as monoidal structure.  This will enable us
to make a qualitative comparison between quantum and classical process networks via the ``descent''

\medskip\centerline{quantum$(\mbox{\bf FDVec},\otimes)\mapsto$ pseudo-quantum$(\mbox{\bf Rel},\times)\mapsto$
classical$(\mbox{\bf Rel},+)$.} 

\medskip 
From a mathematical perspective, part of the formal
investigation consists of ``how much can be done when only
using projectors on a Hilbert space'', which physically
translates as ``how much can be done in terms of state
update due to measurements performed on quantum systems''. 
Although considering state update as a truly dynamical
process introduces an uncertainty on the actual realization
of these processes, it allows a new spectrum of applications
since projections are \emph{not} isomorphisms of the Hilbert
space projection lattice --- by contrast to unitary
transformations which are isomorphisms of that structure.
Furthermore,  in this paper we also prove that  any linear
map on a Hilbert space, including of course  unitary
transformations, can be realized in terms of projections. 

The whole development reveals that projections, or in
physical terms, quantum state update due to measurement,
provide a  powerful tool for high-level approaches to
quantum information processing. In this  context, we also
mention \cite{BIP}.  Thus,
in comparison with the current paradigm with respect to
quantum computation,  involving a setup consisting of a
preparation, a unitary operation and a measurement, we take
a radically different perspective.  In view of the fact that
most of the power of quantum computation with respect to the
known algorithms exploits quantum entanglement, we also mention
that in our setting primitive operations are
``specifications of entanglement'', subject to a Linear
Logical type system.  

The primary technical contributions of this paper are presented in Sections 2.2 and
2.3. In Section 2.1 we recall some basics on vector spaces.  Since in the finite-dimensional case it is
harmless to assume that each complex vector space is in fact a Hilbert space, they provide the approriate
setting for elementary superselection free quantum theory, to which we provide a brief introduction in Section
3.1. The primary conceptual results are presented in Section 3.2 and 3.3. In Section 3.4 and 4 we discuss possible
applications of our results.
 
\section{Constructions for finite-dimensional vector spaces}
  
First we recall some basics on finite-dimensional vector spaces and
establish some notation that
we will need in this paper.  Then we derive the two key lemmas for this paper.
 
\subsection{FD vector spaces and projectors}\label{vectorspaces} 
 
Recall that a finite-dimensional Hilbert space is a complex vector space $({\cal H}, 
{\bf 0}, \cdot , +,{{\mathbb C}})$ equipped with an inner product $\langle-|-\rangle:{\cal
H}\times{\cal H}\to{{\mathbb C}}$ that satisfies 
$$
\begin{array}{ccc} 
&\langle w|c_1\cdot v_1+c_2\cdot v_2\rangle=c_1\langle w|v_1\rangle+c_2\langle w| v_2\rangle\,,&\\ 
&\langle c_1\cdot w_1+c_2\cdot w_2| v\rangle=
\overline{c}_1\langle w_1|v\rangle+\overline{c}_2\langle w_2|v\rangle\,,&\\
&\overline{\langle w| v\rangle}=\langle v|w\rangle\ ,\ 
\langle w| w\rangle=0\,\Rightarrow\, w={\bf 0}\ \ {\rm and}\ \
\langle w| w\rangle\geq 0\,,&
\end{array}
$$ 
where the latter allows us to define a norm on ${\cal H}$ as 
$|-|:=\sqrt{\langle-|-\rangle}$\,.  We introduce an orthogonality relation
$\perp\,\,\subseteq{\cal V}\times{\cal V}$ such that $v\perp w\Leftrightarrow\langle v|w\rangle=0$ and given a
subspace
$V$ of ${\cal V}$ its orthocomplement is 
$$V^\perp:=\{w\in{\cal V}\mid\forall v\in V:v\perp w\}\,.$$
Every finite-dimensional complex vector space extends to a 
Hilbert space via choice of an inner product. 
An orthonormal base is a set of vectors $\{v_i\}_i$ such that
$\langle v_i|v_j\rangle=\delta_{ij}$
where $\delta_{ij}$ is the Kronecker delta, that is, 
$$\delta_{ij}=1\ \ {\rm when}\ \ i=j\ \ {\rm and}\ \
\delta_{ij}=0\ \ {\rm when}\ \ i\not=j\,.$$  
 
A {\it projector} is an idempotent self-adjoint linear endomap $P:{\cal V}\to{\cal V}$ on a Hilbert space ${\cal
V}$, that is respectively, for
$v,w\in{\cal V}$,
$$P(P(v))=P(v)\ \ {\rm and}\ \ \langle P(v)|w\rangle=\langle v|P(w)\rangle\,.$$   
As an example, given a unit vector $v\in{\cal V}$\,, that is
$|v|=1$\,, the map 
$$P_v:{\cal V}\to{\cal V}:w\mapsto\langle v|w\rangle\cdot v$$  
defines a projector.   
The fixpoints of 
$P_v$ constitute the one-dimensional subspace spanned by $v$\,.  

The functionals ${\cal V}^*$\,, i.e. the linear maps $f:{\cal
V}\to{{\mathbb C}}$\,, constitute a vector space isomorphic to ${\cal V}$.  However, there is in general no
canonical isomorphism that connects them. Indeed, since we have
$\langle c\cdot v|-\rangle=\overline{c}\langle v|-\rangle$ the canonical correspondence is anti-linear instead of
linear.  Thus, given a base $\{v_i\}_i$ of ${\cal V}$\,, specification of an isomorphism as ${\cal V}\to{\cal
V}^*:v_i\mapsto\langle v_i|-\rangle$ depends on the choice of the base.  
A functional $\langle
v|-\rangle:{\cal V}\to{{\mathbb C}}$ defines a projector
$\langle{v\over|v|}|-\rangle\cdot{v\over|v|}:{\cal V}\to{\cal V}$ via composition with the injection
${{\mathbb C}}\to{\cal V}:c\mapsto {c\over|v|^2}\cdot v$\,.\footnote{The correspondence between projectors on
one-dimensional subspaces of a finite-dimensional complex vector space and the vector space itself is actually
rather one between the projective space of subspaces, since its points are exactly the one-dimensional
subspaces of the vector space.  More generally,  the (complete) lattice of all subspaces of the
finite-dimensional complex vector space is isomorphic to that of its projectors ordered by $P\leq
Q\Leftrightarrow P;Q=P$. This matter becomes crucial in the passage to quantum theory where states correspond
with one-dimensional subspaces.} For a linear map $f:{\cal V}\to{\cal W}$ and orthonormal bases 
$\{v_i\}_i$ and $\{w_j\}_j$ we have 
$$f=\sum_if_{ij}\langle v_i|-\rangle\cdot w_j\ \ {\rm given\ that}\ \ f(v_i)=f_{ij}\cdot w_j\,.$$    
   
Any pair of complex vector spaces 
${\cal V}$ and ${\cal W}$ admits a tensor product, that is, a pair consisting of a vector space ${\cal
V}\otimes{\cal W}$ and a bilinear map $h:{\cal
V}\times{\cal W}\to{\cal V}\otimes{\cal W}$ such that for any other
bilinear map
$f:{\cal V}\times{\cal W}\to{\cal U}$ there exists a unique $g:{\cal V}\otimes{\cal
W}\to{\cal U}$ with $f=h\,;g$\,. This tensor product equips the category of finite-dimensional complex vector
spaces and linear maps with a monoidal structure with ${{\mathbb C}}^{(1)}$ as unit, since given
$$h:{\cal V}\times{\cal W}\to{\cal V}\otimes{\cal W}\ \ {\rm and}\ \ h':{\cal
V}'\times{\cal W}'\to{\cal V}'\otimes{\cal W}'\,,$$
and two linear maps $f:{\cal V}\to{\cal V}'$ and  
$g:{\cal W}\to{\cal W}'$\,, their tensor product 
$$f\otimes g:{\cal V}\otimes{\cal W}\to{\cal V}'\otimes{\cal W}'$$ 
is uniquely defined due to universality of $h$ with respect to $(f\times g);h'$\,.   

We can construct a tensor product for vector spaces
${\cal V}$ and
${\cal W}$ as 
$h:(v_i,w_j)\mapsto v_i\otimes w_j$ with 
${\cal V}\otimes{\cal W}$ being the vector space spanned by ${\{v_i\otimes w_j\}_{i,j}}$.  
Identifying 
$$(\sum_ic_i\cdot v_i)\otimes(\sum_id_i\cdot w_i)\ \ {\rm and}\ \ 
\sum_{i,j}\!c_id_j\cdot(v_i\otimes w_j )\,,$$ 
this construction does not depend on the choice of
orthonormal base, in particular, for arbitrary $v\in{\cal V}$ and $w\in{\cal W}$ we have $h(v,w)=v\otimes w$\,.
We can define an inner product on ${\cal V}\otimes{\cal W}$ via 
$$\langle v\otimes v'|w\otimes
w'\rangle:=\langle v|w\rangle\langle v'|w'\rangle\,,$$  
so 
$$\Bigl\langle\sum_{i,j}c_{ij}\cdot (v_i\otimes
w_j)\Bigm|\sum_{k,l}c_{kl}\cdot (v_k'\otimes w_l')\Bigr\rangle= 
\sum_{i,j,k,l}\overline{c}_{ij}c_{kl}\langle v_i|v_k'\rangle\langle  
w_j|w_l'\rangle\,.$$  
When both
$\{v_i\}_i$ and $\{w_j\}_j$ are orthonormal bases then   
$\{v_i\otimes w_j\}_{i,j}$ is again orthonormal with respect to this inner product.    
 
The general form 
of elements of ${\cal V}^*\otimes{\cal W}$ and ${\rm Hom}({\cal V},{\cal W})$ respectively is
$$\sum_{i,j}\!c_{ij}\cdot \bigl(\langle v_i|-\rangle\otimes w_j\bigr)\ \ {\rm and}\ \ \sum_ic_{ij}\langle
v_i|-\rangle \cdot w_j$$ and thus we obtain an
isomorphism of vector spaces when providing the set
${\rm Hom}({\cal V},{\cal W})$ with its canonical vector space structure.
Note also that we have ${\cal V}^*\otimes{\cal W}^*\cong({\cal V}\otimes{\cal W})^*$
via identification of $f_i\otimes g_j$\,, with $\{f_i:{\cal V}\to{{\mathbb C}}\}_i$ and $\{g_j:{\cal
V}\to{{\mathbb C}}\}_j$ respective bases for ${\cal V}^*$ and ${\cal W}^*$\,, and the unique functional
$f_i*g_j:{\cal V}\otimes{\cal W}\to{{\mathbb C}}$ that arizes due to universality of $h$ within
$$
\diagram
{\cal V}\otimes{\cal W}&\rTo^{f_i*g_j}&{{\mathbb C}}\\
\uTo^h&\ruTo(2,2)_{f_ig_j}&\\
{\cal V}\times{\cal W}&&\\
\enddiagram
$$
where 
$$f_ig_j:{\cal V}\times{\cal
W}\to{{\mathbb C}}:(v,w)\mapsto f_i(v)g_j(w)\,.$$
The definition for an inner product on ${\cal V}\otimes{\cal W}$ then embodies this fact as
$\langle v\otimes w|-\otimes -\rangle:=\langle v|-\rangle\langle w|-\rangle$
when expressing functionals in terms of
the inner product.   

Categorically, the correspondence between linear maps and the tensor product
arises due to the fact that the category of finite-dimensional
complex vector spaces and linear maps $\mbox{\bf FDVec}$ is compact closed plus the observation that the
internal homs of the form
$[{\cal V},{{\mathbb C}}]$ exactly define the vector space of functionals, since
the dualizing object of the $^*$-autonomous structure and the monoidal unit ${{\mathbb C}}$ coincide,
i.e.~${{\mathbb C}}\cong{{\mathbb C}}^*$\,.  Thus,      
$$[{\cal V},{\cal W}]=[{\cal V},[{\cal W}^*,{{\mathbb C}}]]\cong[{\cal V}\otimes{\cal
W}^*,{{\mathbb C}}]=({\cal V}\otimes{\cal W}^*)^*\cong{\cal V}^*\otimes{\cal W}\,,$$
the last isomorphism
being compact closedness for a $^*$-autonomous category. 
   
\subsection{Implementing the $\mbox{\bf FDVec}$ trace via projectors}  
 
Let $\{e_i\}_i$ be a base of a vector space ${\cal U}$ and let $\{\overline{e}_i\}_i$ be the corresponding
linear functionals in ${\cal U}^*$ via anti-linear correspondence $e_i\mapsto\overline{e}_i$. More
generally, denote by 
$\overline{u}$ the vector $\langle u|-\rangle\in{\cal U}^*$ that corresponds with $u\in{\cal U}$. Let 
$$ 
P_{{\cal U}^*}:=P_{{1\over\sqrt{N}}\cdot\sum_\alpha e_\alpha\otimes \overline{e}_\alpha}:{\cal U}\otimes {\cal
U}^*\to {\cal U}\otimes {\cal U}^*\,,$$
that is 
$$P_{{\cal U}^*}(v)={1\over{N}}\Bigl\langle\sum_\alpha e_\alpha\otimes
\overline{e}_\alpha\Bigm|\,v\Bigr\rangle\cdot\sum_\alpha e_\alpha\otimes \overline{e}_\alpha\,, 
$$
where $N$ denotes the dimension of ${\cal U}$ so that  
$|{1\over\sqrt{N}}\cdot\sum_\alpha e_\alpha\otimes\overline{e}_\alpha|=1$\,.  

Considering the base vectors $\{\overline{e}_i\}_i$ in ${\cal U}^*$ at the right side of the tensor rather than
the base $\{e_i\}_i$  makes the vector $\sum_\alpha
e_\alpha\otimes\overline{e}_\alpha$ base independent, that is, this vector has the same coordinates with
respect to any base of the form $\{e'_i\otimes \overline{e}'_j\}_{i,j}$ of ${\cal U}\otimes{\cal U}^*$\,.  
This motivates the notation $P_{{\cal U}^*}$. (the $^*$
will become clear further)

We can realize the $\mbox{\bf FDVec}$ traces by means of projectors as follows. 
\blm
For any  
$f:{\cal V}\otimes {\cal U}\to{\cal W}\otimes {\cal U}$ and $v\in{\cal V}$ we have for the map 
$$\tau = N\cdot\bigl(\,f\otimes
id_{{\cal U}^*}\ ;\ id_{\cal W}\otimes P_{{\cal U}^*}\,\bigr)\,:\,{\cal V}\otimes
{\cal U}\otimes {\cal U}^*\to {\cal W}\otimes {\cal U}\otimes {\cal U}^*$$
that 
\beq\label{lemmaeq1}  
\tau \Biggl(v\otimes\Bigl(\sum_\alpha {e}_\alpha\otimes
\overline{e}_\alpha\Bigr)\Biggr)\ = \ Tr_{{\cal V},{\cal W}}^{\cal U}(f)(v)\otimes\Bigl(\sum_\alpha
e_\alpha\otimes
\overline{e}_\alpha\Bigr)\,.
\eeq 
Equivalently, setting  
$$\theta = 
(\,id_{\cal W}\otimes P_{{\cal U}^*}\ ;\ \tau\,)
\,:\,{\cal V}\otimes
{\cal U}\otimes {\cal U}^*\to {\cal W}\otimes {\cal U}\otimes {\cal U}^*$$
we have for $u\in{\cal U}\otimes{\cal U}^*$ that
\beq\label{lemmaeq2}
\theta (v\otimes u) \ = \ Tr_{{\cal V},{\cal W}}^{\cal
U}(f)(v)\otimes P_{{\cal U}^*}(u)\,.
\eeq
\elm 
{\bf Proof}. Since we have 
$$(f\otimes id_{{\cal U}^*})\Biggl(v_i\otimes\Bigl(\sum_\alpha {e}_\alpha\otimes 
\overline{e}_\alpha\Bigr)\Biggr)=\sum_{k,l,\alpha}f_{i\alpha{kl}}\cdot (w_k\otimes 
e_l\otimes \overline{e}_\alpha)$$
it follows that 
\begin{eqnarray*}
\tau\Bigl(v_i\otimes\sum_\alpha {e}_\alpha\otimes
\overline{e}_\alpha\Bigr)\hspace{-2.5cm}\\
&=&N\cdot\sum_{k,l,\alpha}f_{i\alpha{kl}}\cdot\Bigl(w_k\otimes
P_{{\cal U}^*}({e}_l\otimes \overline{e}_\alpha)\Bigr)\\   
&=&\sum_{k,l,\alpha}f_{i\alpha{kl}}\cdot\Biggl(w_k\otimes 
\Bigl\langle\sum_\gamma {e}_\gamma\otimes
\overline{e}_\gamma\Bigm|e_l\otimes \overline{e}_\alpha\Bigr\rangle\cdot\Bigl(\sum_\beta   
e_\beta\otimes \overline{e}_\beta\Bigr)\Biggr)\\  
&=&\sum_{k,l,\alpha}f_{i\alpha{kl}}\cdot\Biggl(w_k\otimes 
\Bigl(\sum_\gamma\delta_{\gamma l}\delta_{\gamma\alpha}\Bigr)\cdot\Bigl(\sum_\beta 
e_\beta\otimes \overline{e}_\beta\Bigr)\Biggr)\\
&=&\Bigr(\sum_{k,l,\alpha}\delta_{\alpha l}f_{i\alpha{kl}}\cdot w_k\Bigr)\otimes  
\Bigl(\sum_\beta e_\beta\otimes \overline{e}_\beta\Bigr)\\
&=&Tr_{{\cal V},{\cal W}}^{\cal U}(f)(v_i)\otimes
\Bigl(\sum_\alpha e_\alpha\otimes \overline{e}_\alpha\Bigr)\,.  
\end{eqnarray*}
By linearity
eq.(\ref{lemmaeq1}) then follows. 
We obtain the same result by considering the unit vector ${1\over\sqrt{N}}\cdot\sum_\alpha e_\alpha\otimes 
\overline{e}_\alpha$ in stead of $\sum_\alpha e_\alpha\otimes \overline{e}_\alpha$ in ${v_i\otimes(\sum_\alpha
e_\alpha\otimes \overline{e}_\alpha)}$\,, or more generally, by considering any element of the subspace spanned
by the unit vector
${1\over\sqrt{N}}\cdot\sum_\alpha e_\alpha\otimes 
\overline{e}_\alpha$\,, that is, any element in the image of
$P_{{\cal U}^*}$.  So for arbitrary
$u\in{\cal U}\otimes{\cal U}^*$ eq.(\ref{lemmaeq2}) follows.
\hfill\endproof\bigskip\newline 
Thus the map 
$$
\tau \Biggl(-\otimes\Bigl(\sum_\alpha {e}_\alpha\otimes
\overline{e}_\alpha\Bigr)\Biggr)\,:\,{\cal V}\to {\cal W}\otimes {\cal U}\otimes {\cal U}^*$$
produces $Tr_{{\cal V},{\cal W}}^{\cal
U}(f)(-)$ as the first component of a pure tensor with ${\sum_\alpha
e_\alpha\otimes
\overline{e}_\alpha}$ as {\it context\,} (= the remaining component). Since we
have 
$$v\otimes w={\bf 0}\ \Leftrightarrow\ v={\bf 0}\ {\rm or}\ w={\bf 0}$$ 
the trace is also produced by the map
$$\theta(-\otimes u)\,:\,{\cal V}\to {\cal W}\otimes {\cal U}\otimes {\cal U}^*$$ 
as the first component of a pure tensor with $P_{{\cal U}^*}(u)$ as context provided $P_{{\cal
U}^*}(u)\not={\bf 0}$\,, that is, $u\not\perp\sum_\alpha e_\alpha\otimes \overline{e}_\alpha$ (including
the case $u={\bf 0}$). The function $Tr_{{\cal V},{\cal W}}^{\cal
U}(f)(-)$ is encoded up to rescaling and a
phase factor since in general $(c\cdot v)\otimes w=v\otimes (c\cdot w)$.

Abstracting over the dimension $N$ we can represent eq.(\ref{lemmaeq2})
graphically as      
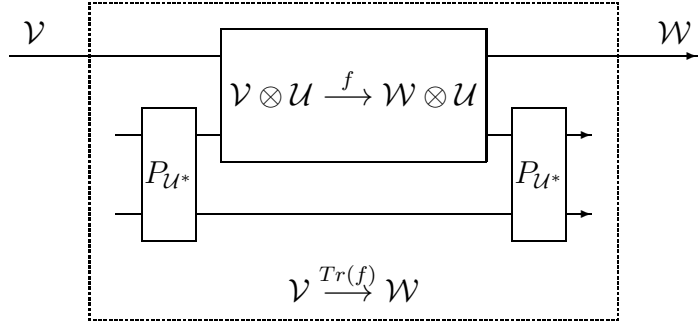
\begin{center}
\begin{picture}(160,120)  

\put(30,50){\framebox(100,50)
{${\cal V}\otimes{\cal U}\stackrel{f}{\longrightarrow}{\cal W}\otimes{\cal U}\vspace{2mm}$}}  
\put(-20,-10){\dashbox(200,120)
{\vspace{-3.2cm}${\cal V}\stackrel{Tr(f)}{\longrightarrow}{\cal W}$}}  
\put(0,20){\framebox(20,50)
{$P_{{\cal U}^*}$}}  
\put(140,20){\framebox(20,50)
{$P_{{\cal U}^*}$}}  
\put(30,90){\line(-1,0){80}} 
\put(130,90){\vector(1,0){80}} 
\put(30,60){\line(-1,0){10}} 
\put(130,60){\line(1,0){10}} 
\put(0,60){\line(-1,0){10}} 
\put(160,60){\vector(1,0){10}} 
\put(0,30){\line(-1,0){10}} 
\put(160,30){\vector(1,0){10}} 
\put(20,30){\line(1,0){120}} 
\put(-45,95){${\cal V}$}
\put(195,95){${\cal W}$}
\end{picture}
\end{center}
\smallskip
\par\bigskip\noindent 
where the arrows that start and end within the dotted lines embody the above mentioned contexts.

Incorporating the dimension $N$ in
order to obtain projections seems an unnecessary complication and indeed, from a purely mathematical 
perspective it is.  However, expressing things in terms of true projectors will be
crucial to us when considering a physical realisation of traces of vector spaces in the next section.  

The fact that we have a traced monoidal category assures
embedding in a compact closed category via the geometry of interaction construction. One verifies that composition
via interaction for $\tilde{f}:{\cal V}_+\otimes{\cal U}_-\to{\cal V}_-\otimes{\cal U}_+$ and 
$\tilde{g}:{\cal U}_+\otimes{\cal W}_-\to{\cal U}_-\otimes{\cal W}_+$ has the following interpretation/realization
(via replacing the trace by its above explicit interpretation/realization)
$$
id_{{\cal V}_+}\!\otimes P_{{\cal U}_-^*}\!\otimes P_{{\cal U}_+}\!\otimes id_{{\cal W}_-}\ ;\
\tilde{f}\otimes id_{{\cal U}^{*}_-}\!\otimes id_{{\cal U}^{*}_+}\!\otimes\tilde{g}\ ;\
id_{{\cal V}_-}\!\otimes (P_{{\cal U}_+^*}\otimes P_{{\cal U}_-})^{\sigma(2,3)}\otimes id_{{\cal W}_+}
$$
where by omitting $^*$ in $P_{{\cal U}_+}$ and $P_{{\cal U}_-}$ we refer to the fact that the two
components in the tensor are swapped, that is 
$$P_{\cal U}:=P_{{1\over\sqrt{N}}\cdot\sum_\alpha \overline{e}_\alpha\otimes
e_\alpha}:{\cal U}^*\otimes {\cal U}\to {\cal U}^*\otimes {\cal U}\,,$$ 
and by ${\sigma(2,3)}$ we mean that the second and the third component in the tensor are
swapped. We then envision this operator as acting on $v_+\otimes u\otimes w_-$ where $u\in{\cal
U}_-\otimes{\cal U}^{*}_-\otimes{\cal U}^{*}_+\otimes{\cal U}_{+}$. Graphically this represents as  
\par
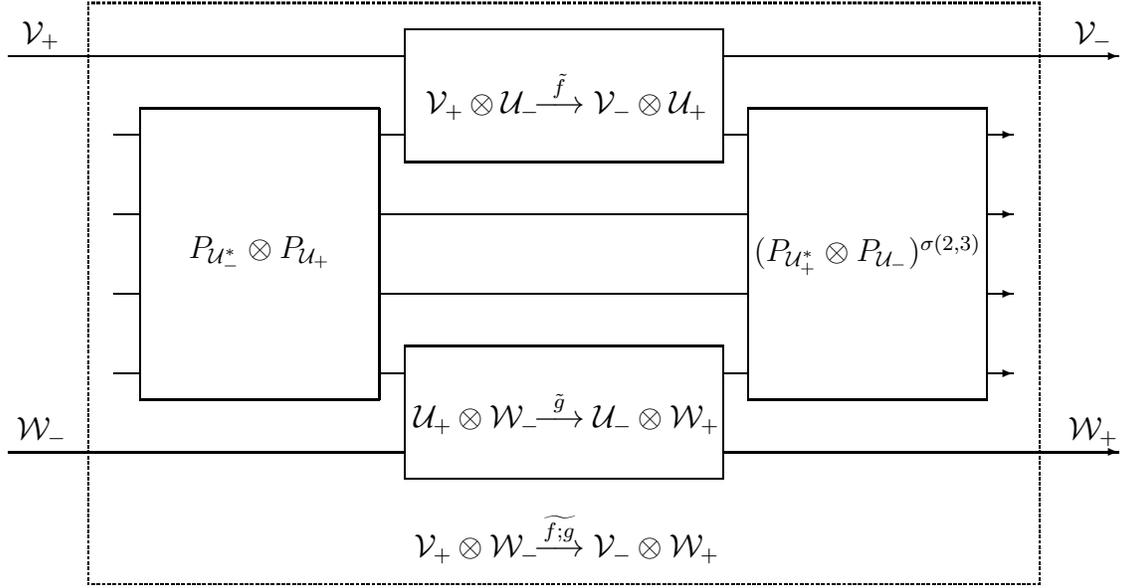
\vspace{-3.5cm}\noindent
\begin{center}
\begin{picture}(180,220)     
\put(30,50){\framebox(120,50)
{${\cal V}_+\hspace{-0.2mm}\otimes{\cal U}_-\hspace{-2mm}\stackrel{\tilde{f}}{\longrightarrow}{\cal
V}_-\hspace{-0.2mm}\otimes{\cal U}_+\hspace{-0.8mm}$}} 
\put(-90,-110){\dashbox(360,220) 
{\vspace{-6.5cm}${\cal V}_+\hspace{-0.2mm}\otimes{\cal W}_-\hspace{-2mm}
\stackrel{\,\widetilde{f;g\,}}{\longrightarrow}{\cal V}_-\hspace{-0.2mm}\otimes{\cal
W}_+\hspace{-0.8mm}$}}   
\put(30,-70){\framebox(120,50)
{${\cal U}_+\hspace{-0.2mm}\otimes{\cal W}_-\hspace{-2mm}\stackrel{\tilde{g}}{\longrightarrow}{\cal
U}_-\hspace{-0.2mm}\otimes{\cal W}_+\hspace{-0.8mm}$}}  
\put(-70,-40){\framebox(90,110)
{$P_{{\cal U}_-^*}\otimes P_{{\cal U}_+}$}}      
\put(160,-40){\framebox(90,110)
{$(P_{{\cal U}_+^*}\otimes P_{{\cal U}_-})^{\sigma(2,3)}$}}  
\put(30,90){\line(-1,0){150}} 
\put(150,90){\vector(1,0){150}} 
\put(30,60){\line(-1,0){10}} 
\put(150,60){\line(1,0){10}} 
\put(-70,60){\line(-1,0){10}} 
\put(250,60){\vector(1,0){10}} 
\put(-70,30){\line(-1,0){10}} 
\put(250,30){\vector(1,0){10}} 
\put(20,30){\line(1,0){140}} 
\put(-70,0){\line(-1,0){10}} 
\put(250,0){\vector(1,0){10}}  
\put(20,0){\line(1,0){140}} 
\put(30,-30){\line(-1,0){10}} 
\put(150,-30){\line(1,0){10}} 
\put(-70,-30){\line(-1,0){10}} 
\put(250,-30){\vector(1,0){10}} 
\put(30,-60){\line(-1,0){150}} 
\put(150,-60){\vector(1,0){150}} 
\put(-115,95){${\cal V}_+$}
\put(283,95){${\cal V}_-$}
\put(-117,-55){${\cal W}_-$}
\put(281,-55){${\cal W}_+$}
\end{picture}
\end{center}
\par\vspace{3.8cm}\noindent
\par\bigskip\noindent
In view of $\mbox{\bf FDVec}$ being compact closed, that is, its trace can be defined as 
$$ 
{\cal V}\stackrel{id_{\cal V}\otimes\eta_{\cal U}}{\longrightarrow}{\cal V}\otimes {\cal U}\otimes
{\cal U}^*\stackrel{f\otimes id_{{\cal U}^*}}{\longrightarrow}{\cal W}\otimes {\cal U}\otimes {\cal
U}^*
\stackrel{id_{\cal W}\otimes\sigma}{\longrightarrow}{\cal W}\otimes {\cal U}^*\otimes
{\cal U}\stackrel{id_{\cal W}\otimes\epsilon_{\cal U}}{\longrightarrow}{\cal W}
$$ 
where corresponding unit and counit are 
$$
\eta_{\cal U}:{{\mathbb C}}\to {\cal U}\otimes{\cal U}^*:1\mapsto\sum_\alpha e_\alpha\otimes
\overline{e}_\alpha
\quad\ \
\epsilon_{\cal U}:{\cal U}\otimes{\cal U}^*\to{{\mathbb C}}:(v,\overline{w})\mapsto \langle w|v\rangle\,,
$$
the crucial part of this construction boils down to the fact that the counit can be interpreted as
$$\epsilon_{\cal U}(e_i\otimes\overline{e}_j)=\langle e_j|e_i\rangle=\delta_{ij}
=\sum_\alpha\delta_{\alpha i}\delta_{\alpha j} =\Bigl\langle\sum_\alpha
{e}_\alpha\otimes\overline{e}_\alpha\Bigm|e_i\otimes
\overline{e}_j\Bigr\rangle$$ or equivalently, $$\epsilon_{\cal U}\otimes\sum_\alpha 
e_\alpha\otimes \overline{e}_\alpha=\Bigl\langle\sum_\alpha
{e}_\alpha\otimes\overline{e}_\alpha\Bigm|-\Bigr\rangle\cdot\sum_\alpha e_\alpha\otimes
\overline{e}_\alpha=N\cdot P_{{\cal U}^*}\,.$$

\subsection{Implementing linear functions via projectors}
    
Given a linear map $f:{\cal V}\to {\cal W}$ and orthonormal bases $\{v_i\}_i$ and $\{w_j\}_j$
set
$$P_f:=P_{{1\over\sqrt{M}}\cdot\sum_{i,j}f_{ij}\cdot(\overline{v}_i\otimes w_j)}:{\cal V}^*\otimes {\cal
W}\to{\cal V}^*\otimes {\cal W}\,,$$
that is 
$$P_f(v)={1\over{M}}\Bigl\langle\sum_{i,j}f_{ij}\cdot(\overline{v}_i\otimes
w_j)\Bigm|\,v\Bigr\rangle\cdot\sum_{i,j}f_{ij}\cdot(\overline{v}_i\otimes w_j)\,, 
$$
where $M=|\sum_{i,j}f_{ij}\cdot(\overline{v}_i\otimes w_j)|^2$.   

The notation $P_f$ is justified by base independence of
${1\over\sqrt{M}}\cdot\sum_{i,j}f_{ij}\cdot(\overline{v}_i\otimes w_j)$ due to the canonical
correspondence between ${\cal V}^*\otimes{\cal W}$ and $[{\cal V},{\cal W}]$.  In fact, the projector
$P_{{\cal U}}$ embodies a particular case of this for $f:=id_{{\cal U}}:{\cal U}\to {\cal U}$ (and
$P_{{\cal U}^*}$ for $f:=id_{{\cal U}^*}:{\cal U}^*\to {\cal U}^*$) given that 
$$ 
id_{\cal U}=\sum_{i,j}\delta_{ij}\langle{e}_i|-\rangle\cdot e_j\ \ {\rm and}\ \ 
N=|\sum_{i,j}\delta_{ij}\cdot(\overline{e}_i\otimes e_j)|^2   
$$
where obviously 
$$\sum_{i,j}\delta_{ij}\cdot(\overline{e}_i\otimes
e_j)=\sum_{\alpha}\overline{e}_\alpha\otimes e_\alpha\,.$$ 

It is our aim to produce arbitrary linear maps using only projectors.
\blm
For any  
$f:{\cal V}\to{\cal W}$ and $v\in{\cal V}$ we have for the map 
$$\xi = N\cdot(P_{{\cal V}^*}\otimes id_{\cal  W})\,:\,{\cal V}\otimes {\cal V}^*\otimes {\cal W}\to{\cal
V}\otimes {\cal V}^*\otimes {\cal W}$$
that
\beq\label{lemmaeq3} 
\xi \Biggl(v\otimes\Bigl(\sum_{i,j}f_{ij}\cdot(\overline{v}_i\otimes w_j)\Bigr)\Biggr)\ = \ \Bigl(\sum_\alpha   
v_\alpha\otimes \overline{v}_\alpha\Bigr)\otimes f(v)\,.
\eeq
Equivalently, setting
$$
\zeta=(\,id_{\cal V}\otimes P_{f}\ ;\ \xi\,)\,:\,{\cal V}\otimes {\cal V}^*\otimes {\cal
W}\to{\cal V}\otimes {\cal V}^*\otimes {\cal W}
$$
we have for $u\in{\cal V}^*\otimes{\cal W}$ 
\beq\label{lemmaeq4} 
\zeta(v\otimes
u)=K\cdot\Bigm(\sum_\alpha v_\alpha\otimes\overline{v}_\alpha\Bigr)\otimes f(v)
\eeq
where $K\in{{\mathbb C}}$ only depends on $u$ and not on $v$.  
\elm
{\bf Proof.} We have 
\begin{eqnarray*}
\xi\Biggl(v_k\otimes\Bigl(\sum_{i,j}f_{ij}\cdot(\overline{v}_i\otimes w_j)\Bigr)\Biggr)\hspace{-3.8cm}\\
&=&N\cdot\sum_{i,j} P_{{\cal V}^*}(v_k\otimes f_{ij}\cdot\overline{v}_i)\otimes w_j\\
&=&\sum_{i,j,\gamma}
\Biggl(\langle v_\gamma\otimes \overline{v}_\gamma| 
v_k\otimes f_{ij}\cdot\overline{v}_i\rangle\cdot\Bigl(\sum_\alpha
v_\alpha\otimes \overline{v}_\alpha\otimes w_j\Bigr)\Biggr)\\  
&=&\sum_{i,j,\gamma}  
f_{ij}\delta_{\gamma k}\delta_{\gamma i}\cdot\Bigl(\sum_\alpha  
v_\alpha\otimes \overline{v}_\alpha\otimes w_j\Bigr)\\ 
&=&\sum_{j}  
f_{kj}\cdot\Bigl(\sum_\alpha  
v_\alpha\otimes \overline{v}_\alpha\otimes w_j\Bigr)\\ 
&=&\Bigr(\sum_\alpha  
v_\alpha\otimes \overline{v}_\alpha\Bigl)\otimes\sum_{j}
f_{kj}\cdot w_j\,.
\end{eqnarray*}
By linearity eq.(\ref{lemmaeq3}) then follows.
Since moreover 
\begin{eqnarray*}
P_f(u)&=&{1\over{M}}\Bigl\langle\sum_{i,j}f_{ij}\cdot(\overline{v}_i\otimes
w_j)\Bigm|\,u\Bigr\rangle\cdot\sum_{i,j}f_{ij}\cdot(\overline{v}_i\otimes w_j)\\
&=&K\cdot\sum_{i,j}f_{ij}\cdot(\overline{v}_i\otimes w_j)
\end{eqnarray*}
for  
$$
K={\Bigl\langle\sum_{i,j}f_{ij}\cdot(v_i\otimes w_j)\Bigm|\,u\Bigr\rangle\over M}
$$
we obtain eq.(\ref{lemmaeq4}).
\hfill\endproof\bigskip\newline
Thus the map 
$$
\xi \Biggl(-\otimes\Bigl(\sum_{i,j}f_{ij}\cdot(\overline{v}_i\otimes w_j)\Bigr)\Biggr)\,:\,{\cal V}\to{\cal 
V}\otimes {\cal V}^*\otimes {\cal W}
$$
produces $f(-)$ as the last component of a pure tensor.  The map 
$$
\zeta(-\otimes u)\,:\,{\cal V}\to{\cal V}\otimes {\cal V}^*\otimes {\cal W}
$$
does the same 
whenever $u\not\perp\sum_{i,j}f_{ij}\cdot(\overline{v}_i\otimes w_j)$  (that is $K\not=0$).  
The function $f$ is again encoded up to rescaling and a
phase factor.

We can now represent the obtained result graphically as 
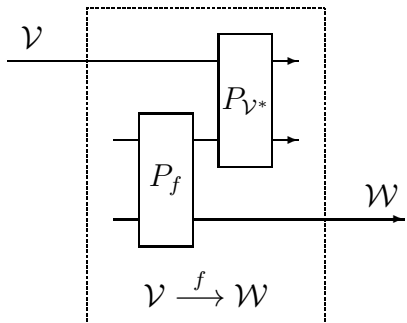
\begin{center}
\begin{picture}(50,123)   

\put(30,50){\framebox(20,50)
{$P_{{\cal V}^*}$}}  
\put(-20,-10){\dashbox(90,120)
{\vspace{-3.2cm}${\cal V}\stackrel{f}{\longrightarrow}{\cal W}$}}  

\put(0,20){\framebox(20,50)
{$P_f$}}

\put(30,90){\line(-1,0){80}} 
\put(50,90){\vector(1,0){10}} 
\put(30,60){\line(-1,0){10}} 
\put(50,60){\vector(1,0){10}} 
\put(0,30){\line(-1,0){10}} 
\put(0,60){\line(-1,0){10}} 
\put(20,30){\vector(1,0){80}}  
\put(-45,95){${\cal V}$}
\put(85,35){${\cal W}$}
\end{picture}
\end{center}
\par\medskip\par\smallskip\noindent 
with the contexts starting and ending within the dotted lines.

\section{Physical realization of $\mbox{\bf FDVec}$ and $\mbox{\bf FRel}$} 
 
We present an  elementary version of quantum theory, restricted to finite-dimensional Hilbert spaces. For a
standard textbook we refer to \cite{Ish} and for more foundationally oriented texts to \cite{vN,Pir,Var}. 
General quantum theory is formulated in terms of infinite-dimensional Hilbert spaces, but for many purposes in
terms of insights, particular quantum features and even structural results, the infinite-dimensional aspects
don't come in, e.g. Gleason's theorem
\cite{Gle} and the Kochen-Specker theorem \cite{KS}, and the finite-dimensional restriction suffices for the
physical realization of the traced monoidal category of finite-dimensional vector spaces.  
 
\subsection{Elementary (superselection free) finite-dimensional quantum theory}  
  
Let ${\cal H}$ be a finite-dimensional (complex) Hilbert space.
Elements of ${\cal H}$ are in the context of quantum theory frequently denoted by $\psi$ and
$\phi$\,. The description of a
quantum system constitutes:
\ben
\item 
Description of the states of the system = kinematics.
\item 
The description of evolution = reversible dynamics.
\item
The description of measurements = non-classical irreversible content.
\een
The states of a quantum system encode as the set 
$\Sigma_{\cal H}$
of one-dimensional subspaces of a Hilbert space ${\cal H}$, structured as an intersection system $(\Sigma_{\cal
H},{\cal L}({\cal H}))$ with ${\cal L}({\cal H})$ the orthocomplemented lattice of closed subspaces of ${\cal H}$ --- 
the classical counterpart to this is the intersection system $(\Sigma,{\cal P}(\Sigma))$ with ${\cal
P}(\Sigma)$ the powerset of the states, that is a complete atomistic Boolean algebra. We motivate this below.
Evolution between time
$t_1$ and $t_2$ is described by a unitary operator, that is, a linear isomorphism that preserves the inner-product
\textit{i.e.} an automorphism of $(\Sigma_{\cal H},{\cal L}({\cal H}))$.

Measurements are represented by self-adjoint operators
$H:{\cal H}\to{\cal H}$. 
When performing a measurement on the system in state
$\psi$ (a notational abuse that we from now on will use freely) where the corresponding self-adjoint operator
has $\sigma(H)$ as its spectrum of eigenvalues  then we obtain as outcome of the measurement a value 
$a\in\sigma(H)$ with corresponding probability
$Prob_\psi^H(a)=\langle\psi|P_a(\psi)\rangle=|P_a(\psi)|^2$\,, where $\psi$ is normalised and $P_a$ is the
projector on the subspace of eigenvectors with eigenvalue $a$.  
Note that $\sum_{a\in\sigma(H)}P_a(\psi)=\psi$ and thus
$\sum_{a\in\sigma(H)}Prob_\psi^H(a)=1$ since all eigenspaces $V_a=\{\psi\in{\cal H}\mid P_a(\psi)=\psi\}$ are
mutually orthogonal and span
${\cal H}$. 
 
Sequential measurements obey von Neumann's  projection postulate \cite{vN}, that is,
if the measurement  yields $a\in\sigma(H)$ as outcome then the state of
the system changes from its initial state $\psi$ to $P_a(\psi)$, so an
immediate next measurement gives again $a$ as outcome --- since $P_a(\psi)$ is itself an 
eigenvector with eigenvalue $a$.     
Thus, projectors encode true state transitions, explicitly 
$$\tilde{P}_a:\Sigma_{\cal H}\setminus {\cal K}\to\Sigma_{\cal H}:\psi\mapsto P_a(\psi)\,,$$ 
where 
$${\cal K}=\{ray(\psi)\in\Sigma_{\cal H}\mid Prob_\psi^H(a)=0\}\,.$$ 

Note that any projector encodes itself a $\{0,1\}$-valued measurement with $V_0\perp V_1$.  Moreover, they can
be envisioned as encoding the primitive propositions on the systems since any self-adjoint operator $H$ on a
finite-dimensional Hilbert space
can always be written as $H=\sum_{a\in\sigma(H)}a
P_a$ --- this is the final
dimensional version of von Neumann's  spectral decomposition theorem \cite{vN}.\footnote{Infinite-dimensionally,
although self-adjoint operators (possibly only partially defined) might have no eigenstates at all, there
always exists a projection valued measure on its spectrum, say $P_{-}^H:{\cal
B}(\sigma(H))\to{{\mathbb P}}({\cal H})$\,, where ${\cal B}(\sigma(H))$ denotes the Borel sets in 
$\sigma(H)$ and ${{\mathbb P}}({\cal H})$ the projectors on ${\cal H}$\,, that
reproduces the self-adjoint operator as $\int_{\sigma(H)}a\, dP^H_a$\,.}   
There is however more. The
primitive propositions for a classical system are the subsets of the state space $\Sigma$. Indeed, let 
$f:\Sigma\to{{\mathbb R}}$ be an observable of a classical system, then
$f^{-1}[E]\in{\cal P}(\Sigma)$ expresses for a state the property 
\begin{quote}
``the value of function $f$ is in $E\subseteq\sigma(f)$''.
\end{quote}
In the quantum
case all statements of the form 
\begin{quote}
``the value of self-adjoint operator $H$ is in $E\subseteq\sigma(H)$''
\end{quote} 
can
be represented by the projector $P_E^H=\sum_{a\in E}P_a^H$, that has $\bigoplus_{a\in E}V_a$ as fixpoints.
Therefore, the subspaces encode the physical properties
attributable to a quantum system.  The projectors can then be envisioned as ``verifications''
\cite{BvN,Pir}, or ``preparations'', or ``{\it active} specifications'' in CS
terms.    

\bigskip\noindent {\bf Example:} {\it Polarization of photons}. Let $Z$ be the 
axis of propagation.  Consider as projector a light analyser that allows only vertically polarized light to pass,
say polarized along the $X$-axis. If the in-coming light is polarized along the $X$-axis it passes
(outcome 1). If the in-coming light is polarized along the $Y$-axis nothing passes (outcome 0).  If it is
polarized along an axis that makes an angle
$\theta$ with the $X$-axis then some light will pass, with relative amplitude ${\rm cos}^2\theta$, and the light
that passed will be vertically polarized.  The amplitude reflects the quantum probability to pass, that is to
obtain an outcome $1$.  The change of polarization angle from $\theta$ to $0$ is then the transition according to
the projection postulate.  The light analyser as such specifies that light is to be vertically polarized and
does this in an active way. 
 
\bigskip   
A quantum system consisting of two subsystems is described in the tensor product ${\cal
H}_1\otimes{\cal H}_2$ where ${\cal
H}_1$ and ${\cal H}_2$ are the Hilbert spaces in which we describe the respective subsystems.
Thus, whereas in
classical physics two systems are described by pairing states --- the Cartesian
product --- in quantum theory we also have to consider superpositions of such pairs. 
Examples of projectors on ${\cal
H}_1\otimes{\cal H}_2$ are those of the form $P_1\otimes P_2$, explicitly definable as
$$(P_1\otimes P_2)(\psi\otimes\phi)=P_1(\psi)\otimes P_2(\phi)\,.$$
Even though values for general self-adjoint operators of this form should be envisioned as pairs
$(a_1,a_2)\in\sigma(H_1)\times\sigma(H_2)$ with corresponding probabilities
$$Prob_\Psi^{H_1\otimes H_2}(a_1,a_2) 
=\langle\Psi|P_{a_1}\otimes P_{a_2}\Psi\rangle\,,$$
projectors compose conjunctively under 
$\otimes$ that is $(0,1)\sim(0,0)\sim(1,0)\sim 0$ and $(1,1)\sim 1$. Other examples of projectors on ${\cal
H}_1\otimes{\cal H}_2$ are
$P_\Psi=\langle\Psi|-\rangle\Psi$ where $\Psi$ cannot be written as a pure tensor.  
The examples in the previous
section were all of this so-called ``non-local'' form.\footnote{`Non-local'
should not, in this context,  be envisioned in space-like terms. Non-local unitary operations are
considered in quantum control theory, quantum computation and quantum information \cite{BCL}, when the system
evolves according to a non-local Hamiltonian, i.e.
$U(t)=e^{{i\over\bar{h}}Ht}\not=U_1(t)\otimes U_2(t)=e^{{i\over\bar{h}}H_1t}\otimes 
e^{{i\over\bar{h}}H_2t}$. They allow to obtain non-local projectors as  
$U.(P_1\otimes P_2).U^{-1}$ when $U\not=U_1\otimes U_2$\,. Any one-dimensional projector $P$ on ${\cal
H}_1\otimes{\cal H}_2$ can be obtained in that way as $U.P_{\psi\otimes\phi}.U^{-1}$ taking $U$ such that
$U^{-1}(\psi\otimes\phi)$ is a fixpoint of $P$ since $P_{\psi\otimes\phi}=P_\psi\otimes P_\phi$\,. Note that
one cannot obtain arbitrary projectors in this way due to the simple fact that the dimension of the projector on
the global space should factor in a product of the dimensions of the underlying ones, e.g a projector on a
$5$-dimensional subspace (with $7$-dimensional orthocomplement) in case of $dim({\cal H}_1)=3$ and $dim({\cal
H}_2)=4$.}   

\subsection{Physical realization of $(\mbox{\bf FDVec},\otimes,Tr)$.}  

The physical realization of $\mbox{\bf FDVec}$ as a traced monoidal category consists of interpreting its
objects, morphisms and additional operations, the tensor bifunctor
and the trace, in physical terms, analogous to the realization of $(\mbox{\bf Rel},+)$ presented in the
introduction in terms of tokens carrying data traveling in a network.  
 
\medskip\noindent
{\bf i. ``$\mbox{\bf FDVec}$-Objects''.}  
${\cal V}\in Ob(\mbox{\bf FDVec})$ is interpreted/realized as a quantum system described in corresponding Hilbert
space.    We can represent this quantum system by its trajectory  
\begin{center}
\begin{picture}(150,18)   
\put(0,0){\vector(1,0){150}}  
\put(60,5){$\psi\in{\cal W}$} 
\end{picture}
\end{center}
where the direction of the arrow should be read as the flow of time.   
 
\medskip\noindent
{\bf ii. ``$\mbox{\bf FDVec}$-Morphisms''.} $f\in{\bf FDVec}({\cal V},{\cal W})$ is interpreted as the
process obtained by interpreting in Lemma 2 the vector spaces as quantum systems and the projectors as
$\{0,1\}$-measurements.\footnote{Since states encode in 
Hilbert space as rays rather than as vectors, two functions $f$ and $g$ which are such that
$f=re^{i\theta}\cdot g$ with $r\in{\mathbb R}_+$ and $\theta\in[0,2\pi[$ will be encoded by
the same physical process.   Due to linearity of the maps however, coincidence of the action of $f$ and
$g$ on one-dimensional subspaces does force them to be essentially the same \cite{FF1,FF2,FF3}. Quantum
mechanics moreover provides sufficient tools to even encode phase factors in a measurable way by
de-localising one part of the state from another part
\cite{Rau}.  Relative amplitudes can be envisioned as relative densities.  But, more importantly, the
minor unfaithfulness, when retained, is not essential at all for the qualitative results we pursue and
does not seem to play any role in any applications we can think of.}    

\medskip\noindent
{\bf iii. ``$\mbox{\bf FDVec}$-Tensor''.} ${\cal V}\otimes{\cal W}\in Ob(\mbox{\bf FDVec})$ is interpreted
as a compound quantum system which extends to 
$f\otimes g\in{\bf FDVec}({\cal V}\otimes{\cal V}',{\cal W}\otimes{\cal W}')$ as non-interacting  parallel
composition.  The network consists of a number of parallel tracks on which quantum particles synchronously
travel ``as a wave'' 
$$
\sum_{i_1,\ldots,i_n}c_{i_1\ldots i_n}\psi_{i_1}\otimes\ldots\otimes\psi_{i_n}, 
$$
being acted on ``locally'' (that is on part of the wave) by processes.

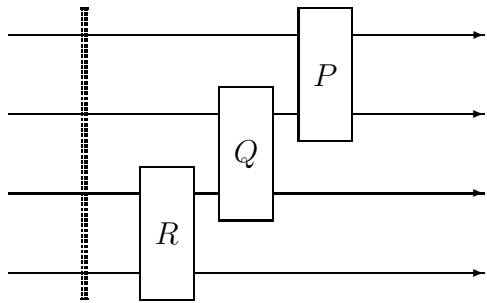
\begin{center}
\begin{picture}(50,100)   

\put(30,50){\framebox(20,50)
{$P$}}  
\put(0,20){\framebox(20,50)
{$Q$}}  
\put(-30,-10){\framebox(20,50)
{$R$}}  

\put(30,60){\line(-1,0){10}} 
\put(0,30){\line(-1,0){10}} 
\put(-30,30){\line(-1,0){50}} 
\put(-30,0){\line(-1,0){50}} 
\put(30,90){\line(-1,0){110}} 
\put(0,60){\line(-1,0){80}} 
\put(50,90){\vector(1,0){50}} 
\put(50,60){\vector(1,0){50}} 
\put(20,30){\vector(1,0){80}}  
\put(-10,0){\vector(1,0){110}}  
\put(-50,-10){\dashbox(0,110)
{\vspace{-3.2cm}\hspace{0cm}}}  
\put(-51,-10){\dashbox(0,110)
{\vspace{-3.2cm}\hspace{0cm}}}  
\put(-52,-10){\dashbox(0,110)
{\vspace{-3.2cm}\hspace{0cm}}}  
\end{picture}
\end{center}

\smallskip\noindent
The dashed line represents a propagating wave front.

A succesful pass of a wave of particles through the
network requires a succesful pass through every projector. Note that the propagation of the wave can as well be
envisioned as being active and passive, that is, either the particles propagate themselves or the processes
acting on them.

Note that in these physical terms Lemma 2 embodies what could be called ``generalised probabilistic
teleportation'', with ``probabilistic quantum teleportation'' and ``probabilistic quantum cloning'' 
as instances.\footnote{The no-cloning theorem for
quantum states prohibits perfect (certain) copying of quantum states by means of `physical operations', usually
envisioned as trace preserving completely positive maps on density matrices --- see for example \cite{WZ} or
\cite{Joz}.  For the standard quantum teleportation protocol we refer to \cite{BBC}.} 
It suffices to set respectively 
$$f=id_{\cal V}:{\cal V}\to{\cal V}:\psi_i\mapsto \psi_i$$ 
and
$$f=\Delta_{\cal V}:{\cal V}\to{\cal V}\otimes{\cal V}:\psi_i\mapsto(\psi_i,\psi_i)$$
with $\{\psi_i\}_i$ a base of ${\cal V}$.

\medskip\noindent
{\bf iv. ``$\mbox{\bf FDVec}$-Trace''.} $Tr_{{\cal V},{\cal W}}^{\cal U}$ for $f\in{\bf FDVec}({\cal
V}\otimes{\cal U},{\cal W}\otimes{\cal U})$ is interpreted as the
process obtained  from Lemma 1 by interpreting the
vector spaces as quantum systems and the projectors as measurements, with again the same remark on the weights $N$ as
above.\footnote{Note here the fundamental difference between the vector space trace with respect to this realisation and
the realization of the partial trace for unitary operations acting on so-called bipartite states (as encountered in
standard quantum information textbooks, e.g. for the realization of superoperators).  Here we sum over pure states
where the usual partial trace realization yields a sum over density operators.  In general, that partial
trace realization also does not de-entangle.} 
   
\medskip\noindent
Since we realized all operations that enable the
GoI-construction, we realized the GoI-construction for $\mbox{\bf FDVec}$ itself.
For the whole interpretation we need only state update (projectors) and no unitary operators,
which are themself morphisms realized in terms of state update.   

Note in particular that all processes in
the network can be reduced (via Lemma 2) to the form $P_{{\cal U}^*}$ or $P_f$, where we even have
$P_{{\cal U}^*}=P_{id_{{\cal U}^*}}$.  
In full notation we as such only used
$$
P_{f\,:{\cal V}\to{\cal W}}\ \ \ {\rm and}\ \  \ P_{id\,:\,{\cal U}\to{\cal U}}\,.
$$
Thus, all processes can be typed by the Linear Logic encoded in
$\mbox{\bf FDVec}$.  

Note here also that the isomorphism $[{\cal V},{\cal W}]\cong{\cal V}^*\otimes{\cal
W}$ is physically exploited as an isomorphism between application of arbitrary linear functions and
projection on corresponding entangled states.
This suggests the slogan that ``basic types of quantum theory are themself
actions'', represented by projections.  The projectors of the
form $P_f$ then encode specification of a particular type of entanglement, where the corresponding action itself is
then the process of entangling. 

\subsection{Physical realization of $(\mbox{\bf FRel},\times,Tr)$}\label{RelReaealization} 

We will now establish a functorial mapping that carries the trace structure of $(\mbox{\bf FDVec},\otimes)$ on
$(\mbox{\bf FRel},\times)$ and hence also the physical realization.
We recall that $\mbox{\bf FRel}$ is equivalently described as the
Kleisli category of the covariant powerset monad on $\mbox{\bf FSet}$,
or equivalently again the category of finitely-generated free
suplattices and their homomorphisms.
Concretely, the isomorphism from $\mbox{\bf FRel}$ to this latter
category is described as follows.
Sets
$X$ in $\mbox{\bf FRel}$ are represented as powersets ${\cal P}(X)$ in ${\rm P}(\mbox{\bf FSet})$ and relations
$R\subseteq X\times Y$ as maps 
$$f_R:{\cal P}(X)\to{\cal P}(Y):T\mapsto\{y\in Y\mid\exists t\in T: (t,y)\in R\}\,.$$   

Assume that every vector space ${\cal V}$ has a specified base $\{e_i\}_{i\in X}$,
explicitly denoted as $({\cal V},\{e_i\}_{i\in X})$, and that for $({\cal V},\{e_i\}_{i\in X})$ and
$({\cal W},\{e_j'\}_{j\in Y})$ the base of ${\cal V}\otimes {\cal W}$ is $\{e_i\otimes e'_j\}_{(i,j)\in X\times
Y}$. Define (with abuse of the notation $\mbox{\bf FDVec}$):
$$
\tilde{\rm R}:\mbox{\bf FDVec}\to\mbox{\bf FRel}:  
\bla
({\cal V},\{e_i\}_{i\in X})\mapsto X\\
f\mapsto
\tilde{\rm R}_f:=\{(i,j)\in X\times Y\mid f_{ij}\not=0\}. 
\ela
$$ 
Since   
\begin{eqnarray*}
\tilde{\rm R}_{f;g}&=&\{(i,k)\in X\times Z\mid \sum_j f_{ij}g_{jk}\not=0\}\\ 
&\subseteq&
\{(i,k)\in X\times Z\mid \exists j\in Y: f_{ij}g_{jk}\not=0\}=\tilde{\rm R}_f;\tilde{\rm R}_g
\end{eqnarray*} 
and 
$$\tilde{\rm R}_{id_X}=\{(i,j)\in X\times  X\mid \delta_{ij}\not=0\}=\Delta_X$$
it follows that $\tilde{\rm R}$ is a lax functor.  To show that it is
not a (strict) functor, it
suffices to consider
$f:{{\mathbb R}}^2\to{{\mathbb R}}^2$ with $f_{ij}={1\over\sqrt{2}}$ except for $f_{22}=-{1\over\sqrt{2}}$.
Then we have
$f;\!f=id_{{{\mathbb R}}^2}$ so clearly ${\rm R}(f;\!f)\not={\rm R}f;{\rm R}f$.  So the presence of negative
values, or more general, complex phases, allows ``mutual cancelation'', a qualitative feature of $\mbox{\bf
FDVec}$ that has no counterpart in $\mbox{\bf FRel}$.  When restricting to those morphisms in
$\mbox{\bf FDVec}$ which only have non-negative real coefficients
$f_{ij}$ we have $\sum_j f_{ij}g_{jk}\not=0 \Leftrightarrow \exists j\in Y: f_{ij}g_{jk}\not=0$ and thus
obtain a true functor.  Denote this category by
$\mbox{\bf FDVec}_+$ and from now on we set
${\rm R}:\mbox{\bf FDVec}_+\to\mbox{\bf FRel}$ for the restriction.   This functor preserves the tensor by
construction and also preserves the trace.    Indeed, 
\begin{eqnarray*}
{\rm R}_{Tr^{\cal U}_{{\cal V},{\cal W}}f}
&=&\{(i,j)\in X\times Y\mid \sum_k f_{ikjk}\not=0\}\\
&=&\{(i,j)\in X\times Y\mid \exists k\in Z: f_{ikjk}\not=0\}
=Tr^{{\rm R}{\cal U}}_{{\rm R}{\cal V},{\rm R}{\cal W}}{\rm R}_f
\end{eqnarray*} 
since 
$${\rm R}_f=\{((i,l),(j,k))\in(X\times Z)\times(Y\times Z)\mid f_{iljk}\not=0\}\,.$$   
 
Concretely, in terms of ${\rm P}(\mbox{\bf FSet})$ rather than $\mbox{\bf FRel}$, the above functor
assigns to $v\in(V,\{e_i\}_{i\in X})$ the set of all $i\in X$ such that $\langle v|e_i\rangle\not=0$, idem ditto
for the images of the base vectors under linear maps, so in particular we have 
$$({\rm R}f)(i)=\{j\in
Y\mid f_{ij}\not=0\}$$ 
and in general for $T\subseteq X$
$$({\rm R}f)(T)=\{j\in Y\mid \exists i\in T:f_{ij}\not=0\}\,.$$ 
 
{\it This concreteness is essential.} It enables us to define a system
with  states given by equivalence classes of states
of a quantum system, which we will denote by elements of ${\cal P}(X)$:
The system is in state $T\in{\cal P}(X)$ if the quantum system is in a state of the form $\sum_{i\in
T}c_i\cdot\psi_i$ with all $c_i$ non-zero.
The physically meaningful operations for the quantum system, in vector space terms composition, tensor and trace,
are all preserved. As such we realize $(\mbox{\bf FRel},\times,Tr)$ physically. 
  
Note that we could not obtain a realization via embedding, since
there exists no functorial embedding  
${\rm F}:{\rm P}(\mbox{\bf FSet})\to\mbox{\bf FDVec}$ that assigns
to each $\{i\}\in {\cal P}(X)$ a corresponding  base vector
$e_i\in\{e_j\}_j$.   This fact reflects a ``resource
sensitivity'' of linear maps, that has no counterpart for relations, in the sense of ``how many 
elements in the argument contribute to a particular element of the image''.  
This is exactly what is
captured by so-called multirelations, that is, the category
$\mbox{\bf Mult}_{{{\mathbb N}}}$ of sets with as morphisms maps
$r:X\times Y\to{{\mathbb N}}:(i,j)\mapsto r_{ij}$.  In the above setting, this category can be made concrete
when restricting to finite sets by envisioning the sets $X$ themself as the collection of maps
$${\cal P}_{{{\mathbb N}}}(X):=\{t:X\to{{\mathbb N}}:i\mapsto t_i\}$$
on which the multirelations act as 
$$f_r:{\cal P}_{{{\mathbb N}}}(X)\to{\cal P}_{{{\mathbb N}}}(Y):t\mapsto
f_r(t):=Y\to{{\mathbb N}}:j\mapsto\sum_{i\in X} t_i\,r_{ij}\,$$  Note that the same construction still holds
when considering ${{{\mathbb R}}_+}$ instead of ${{\mathbb N}}$. Denoting continuously valued 
multirealtions on finite sets as $\mbox{\bf FMult}_{{{\mathbb R}}_+}$ we obtain $\mbox{\bf
FDVec}_+\cong\mbox{\bf FMult}_{{{\mathbb R}}_+}$ which itself is then naturally equipped with tensor and trace. 
 
It is possible to
choose the bases $\{e_i\}_{i\in X}$ such that $\mbox{\bf FDVec}_+$ is closed under the canonical $\mbox{\bf
FDVec}$
$^*$-operation.  It suffices to set $({\cal V}^*,\{\overline{e}_i\}_{i\in X})$ whenever we have
$({\cal V},\{e_i\}_{i\in X})$.  Then we have that $f^*_{ij}=f_{ji}$, as such inheriting being positive reals and
moreover, the induced $\mbox{\bf FDVec}_+$ $^*$-operation is preserved by the functor ${\rm R}$ since the
canonical $^*$-operation in $(\mbox{\bf FRel},\times)$ on objects is
the identity and on morphisms is the relational
converse.
  
\subsection{Delineation of qualitative computational differences}   

The above construction enables to delineate qualitative differences between quantum and
classical process networks, where we envision the latter described by $(\mbox{\bf Rel},+)$ as discussed in
the introduction.  We will proceed via a two step ``descent'' $(\mbox{\bf FDVec},\otimes)\mapsto(\mbox{\bf
Rel},\times)\mapsto(\mbox{\bf Rel},+)$ in terms of their physical realizations. Following section
\ref{RelReaealization}, the passage from $(\mbox{\bf FDVec},\otimes)$ to
$(\mbox{\bf Rel},\times)$ via the functor ${\rm R}$ goes with the following ``loss'':
\bit 
\item Specification of phase-factors due to the domain restriction of
  ${\rm R}$ to $\mbox{\bf FDVec}_+$, corresponding to the loss of the possibility of mutual cancellation.
In view of the importance of ``mutual cancellation'' in known quantum computational algorithms this
feature is essential. Note that this distinction also appears when comparing probabilistic and
quantum Turing machines.
\item Expressibility of multiplicities since ${\rm R}$ is not
  faithful, corresponding to the loss of degrees of freedom
at the level of the system traveling through the network.
\item Processing along multiple incompatible bases becomes impossible,
  corresponding to the loss of degrees of freedom
at the information processing level.
\eit 
Next we discuss the passage from $(\mbox{\bf Rel},\times)$ and $(\mbox{\bf Rel},+)$ where we encounter
two crucial physical differences.
\bit
\item In the case of the realization of $(\mbox{\bf Rel},+)$ multiple images for an initial state
of a process stands for ``non-determinism'', in the case of the realization of $(\mbox{\bf Rel},\times)$ it
stands for ``doing multiple things together'' via creation of a superposition state that reflects all possible
images for a single argument.
\item In the case of the realization of $(\mbox{\bf Rel},+)$ the trace is realized recursively, in the case
of the realization of $(\mbox{\bf Rel},\times)$ it is a one-shot passage, with corresponding implications in
terms of complexity.  
\eit 
 
\section{Applications} 

The present paper, while very much a first step, lays the basis for a
number of further developments.  We realized  wave-style GoI models of Multiplicative 
Linear Logic via the quantum physical processes {\it entangling} and {\it de-entangling} by
means of typed projectors.
In particular, we can now give a `physical realization' of proofs in the
Multiplicative fragment of Linear Logic, or of terms in the
(simply-typed) \emph{affine} $\lambda$-calculus, as quantum systems.
Since e.g. boolean circuits can easily be represented as affine lambda 
terms, this gives us a `compilation process' taking high-level
functional programs into quantum systems, of a form which looks very
different to the current low-level descriptions of quantum algorithms
and machine models. It will surely be of interest to look at examples
of this compilation process and compare them to current approaches. Since there
are currently very few quantum algorithms, it is to be hoped that such 
higher-level methods will be fruitful in suggesting new ideas and approaches. 
We can also pose the following questions for future investigation. 
\bit
\item
Can we realize proof reduction (normalization) in linear $\lambda$-calculus in constant time via physical
processes, due to the one shot trace of the computational model proposed in this paper? 
\item
Can one produce a general picture of wave-style GoI in terms of quantum structures?  A possible candidate is a category
of orthoalgebras, which captures both sets and general quantum
structures, and admits a tensor product. See for example \cite{CMW}.
\eit


\begin{thebibliography}{99}      
\bibitem{Abr} 
Abramsky, S., {\em Retracing some Paths in Process Algebra}, ``Proceedings of 
the Seventh International Conference on Concurrency Theory'' LNCS
{\bf 1119} (1996), 1--17. 
\bibitem{AHS}
Abramsky, S., E. Haghverdi and P. J. Scott, {\em Geometry of Interaction and Linear Combinatory
Algebras}, Mathematical Structures in Computer Science, To appear.
\bibitem{AJ} 
Abramsky, S. and R. Jagadeesan, {\em New Foundations for the Geometry of Interaction}, Information and
Computation {\bf 111} (1994), 53--119.
\bibitem{Bai} 
Bainbridge, E. S., {\it Feedback and Generalized Logic}, Information and Control
{\bf 31} (1976), 75--96.
\bibitem{BBC}
Bennet, C. H., C. Brassard, C. Cr\'epeau, R. Jozsa, A. Peres and
W. K. Wooters, {\em Teleporting an Unknown Quantum State via Dual Classical and Einstein-Podolsky-Rosen
Channels},  Physical Review Letters {\bf 70} (1993), 1895--1899.
\bibitem{BCL} 
Bennett, C. H., J. I. Cirac, D. Leifer, D. W. Leung, N. Linden, S. Poperscu
and G. Vidal, {\em Optimal Simulation of Two-Qubit Hamiltoneans using
General Local Operations},
URL: {\tt http://arXiv.org/abs/quant-ph/ 0107035}.
\bibitem{BvN}
Birkhoff, G. and J. von Neumann, {\em The Logic of
Quantum Mechanics}, Annals of Mathematics {\bf 37} (1936), 823--843.
\bibitem{BIP}
Blute, R. F., I. T. Ivanov and P. Panangaden, {\em Discrete Quantum
Causal Dynamics}, URL: {\tt http://arXiv.org/abs/quant-ph/0109053}.
\bibitem{CMW}
Coecke, B., D. J. Moore and A. Wilce, Eds., ``Current Research in
Operational Quantum Logic: Algebras, Categories, Languages'',
Kluwer Academic Publishers, 2000. 
\bibitem{FF1}
Faure, Cl.-A. and A. Fr\"olicher, {\em Morphisms
of Projective Geometries and of Corresponding Lattices}, Geometriae
Dedicata {\bf 47}, (1993) 25--40.
\bibitem{FF2}
Faure, Cl.-A. and A. Fr\"olicher, {\em Morphisms
of Projective Geometries and Semilinear Maps}, Geometriae
Dedicata {\bf 53}, (1994) 237--262.
\bibitem{FF3}
Faure, Cl.-A. and A. Fr\"olicher, 
``Modern Projective Geometry'',
Kluwer Academic Publishers, 2000.
\bibitem{Joz} 
Jozsa, R., {\it A Stronger No-Cloning Theorem},
URL: {\tt http://arXiv.org/
abs/quant-ph/0204153}.
\bibitem{Gle} 
Gleason, A. M., {\em Measures on the Closed Subspaces of
a Hilbert Space}, Journal of Mathematics and Mechanics {\bf 6} (1957),
885--893.
\bibitem{Hag}
Haghverdi, E., ``A Categorical Approach to Linear Logic, Geometry of Proofs and Full Completeness'', PhD Thesis,
University of Ottawa, 2000.
\bibitem{Ish} 
Isham, C. J., ``Lectures on Quantum Theory'', Imperial College Press, 1995.
\bibitem{JSV}
Joyal, A., R. Street and D. Verity, {\em Traced Monoidal Categories}, Proceedings of the
Cambridge Philosophical Society {\bf 119} (1996), 447--468.
\bibitem{KS}
Kochen, S. and E. P. Specker, {\em The Problem of Hidden Variables in Quantum Mechanics}, 
Journal of Mathematics and Mechanics {\bf 17}, (1967) 59--87.
\bibitem{vN}
von Neumann, J., ``Mathematische Grundlagen der
Quantenmechanik'', Springer--Verlag, 1932. Translation in ``Mathematical Foundations of Quantum
Mechanics'', Princeton University Press, 1955.
\bibitem{Pir}
Piron, C., ``Foundations of Quantum Physics'', W.A. Benjamin, 1976.
\bibitem{Rau}
Rauch, H., W. Treimer and U. Bonse, {\em Test
of a Single Crystal Neutron Interferometer}, Physics Letters
{\bf 47 A}, (1974) 369--384. 
\bibitem{Sel1}
Selinger, P., {\em A Note on Bainbridge's Powerset Construction},
URL: {\tt ftp://quasar.mathstat.uottawa.ca/pub/selinger/bainbridge.ps.gz}.
\bibitem{Sel2}
Selinger, P., {\em Categorical Structure of Asynchrony}, ``Proceedings of MFPS 15'', 
Electronic Notes in Theoretical Computer Science {\bf 20} (1999).
\bibitem{Var} 
Varadarajan, V. S., ``The Geometry of Quantum Theory'', Springer-Verlach, 1968.
\bibitem{WZ}
Wootters, W. and W. Zurek, {\em A Single
Quantum cannot be Cloned}, Nature {\bf 299} (1982), 802--803.  
\end{thebibliography}
\end{document}